\documentclass[aps,twocolumn,showpacs]{revtex4}
\usepackage{bm}
\usepackage{color} 
\usepackage{graphicx}
\usepackage{epsfig}
\usepackage{hyperref}
\usepackage{natbib}

\newcommand{\Tr}{\mathop{\rm Tr}\nolimits}
\newcommand{\trans}[1]{{#1}^{ T}}
\newcommand{\smt}[1]{{#1}^{({ s})}}
\newcommand{\asmt}[1]{{#1}^{({ a})}}
\newcommand{\cen}[1]{{#1}^{({ c})}}
\newcommand{\tG}{\bm{\Gamma}}
\newcommand{\ts}{\bm{\sigma}}
\newcommand{\tL}{\bm{\Lambda}}

\renewcommand{\Re}{{\rm Re\,}}
\renewcommand{\Im}{{\rm Im\,}}

\begin{document}

\title{Electron spin dephasing in two-dimensional systems with anisotropic scattering}

\author{A.\,V.\,Poshakinskiy}
\author{S.\,A.\,Tarasenko}

\affiliation{A.F.~Ioffe Physical-Technical Institute, Russian Academy of Sciences, 194021 St.~Petersburg, Russia}


%
%
%
%
%
%

\begin{abstract}
We develop a microscopic theory of spin relaxation of a two-dimensional electron gas in quantum wells with anisotropic electron scattering. Both precessional and collision-dominated regimes of spin dynamics are studied. It is shown that, in quantum wells with noncentrosymmetric scatterers, the in-plane and out-of-plane spin components are coupled: spin dephasing of carriers initially polarized along the quantum well normal leads to the emergence of an in-plane spin component even in the case of isotropic spin-orbit splitting. In the collision-dominated regime, the spin-relaxation-rate tensor is expressed in terms of the electric conductivity tensor. We also study the effect of an in-plane and out-of-plane external magnetic field on spin dephasing and show that the field dependence of electron spin can be very intricate.         
\end{abstract}

\maketitle

\section{Introduction}

The spin dynamics of charge carriers in semiconductor structures has been attracting a great deal of attention~\cite{Dyakonov08:SPS}. Much effort is focused on experimental and theoretical studying the electron spin dephasing in quantum wells (QWs) and obtaining the controllable spin lifetime (for a recent review see Refs.~\onlinecite{Wu10,Korn10,Glazov10,Muller10}). It is established that, in a wide range of temperature, carrier density and mobility, the spin lifetime of a two-dimensional electron gas is limited by the D'yakonov-Perel' (DP) spin dephasing mechanism~\cite{Dyakonov71,Dyakonov86}. The mechanism is based on precession of individual electron spins in the Rashba and/or Dresselhaus effective magnetic field 
and is highly sensitive to the QW crystallographic orientation~\cite{Dyakonov86,Averkiev99,Cartoixa05,Tarasenko09} as well as 
details of electron scattering by structural defects and phonons. Depending on the ratio between the period of spin precession in the effective field and the momentum relaxation time of carriers, the spin polarization monotonically decays or exhibits damping oscillations~\cite{Gridnev01,Brand02,Leyland07,Griesbeck09}.  
So far, the DP mechanism has been theoretically analyzed for central electron scattering, neglecting possible anisotropy of scattering potential. However, such a model does not always describe electron scattering in QWs adequately. Transport measurements reveal that electron mobility and scattering rate can be anisotropic in the QW plane even for (001)-grown structures~\cite{Papadakis02,Ercolani08,Akabori10}.
Strong in-plane anisotropy of electric properties has been also demonstrated recently for QW structures with embedded semidisk-shaped or elongated dots~\cite{Sassine08,Li11}.   

In the present paper, we study the electron spin dephasing in QW structures with anisotropic scattering potential and derive equations for the spin-relaxation-rate tensor.
We show that anisotropic scattering qualitatively modifies the spin dephasing both in the collision-dominated and oscillatory regimes. The paper is organized as follows. In Sec.~\ref{Sec_gen}, we present a general formalism for describing the electron spin dynamics in quantum wells in the presence of anisotropic elastic scattering. Collision-dominated regime of the DP spin dephasing is considered in Sec.~\ref{Sec_collisions}. We show that 
the spin-relaxation-rate tensor can be expressed in terms of the constants of spin-orbit splitting and the electric conductivity tensor. In (001)-grown QWs with anisotropic scatterers,
the longest lifetime of electron spin along the growth direction is achieved in QWs with structure inversion asymmetry where the Rashba effective field is nonzero. 
The oscillatory regime of spin dephasing, which is realized in high-mobility QWs, is considered in Sec.~\ref{Sec_oscillations}. It is shown that anisotropic scattering leads to a coupling between the in-plane and out-of-plane spin components even in the case of isotropic Rashba or Dresselhaus spin-orbit splitting. In particular, the spin dephasing of carriers initially polarized along the QW normal leads to the emergence of a net in-plane spin component which then also vanishes. We also analyze the effect of an external magnetic field on spin dephasing both in the collision-dominated and oscillatory regimes and show that the field dependence of electron spin can be very intricate. The main results of the paper are summarized in Sec.~\ref{Summary}.

\section{General equations}\label{Sec_gen}

The time evolution of the spin distribution function $\bm{s_k}$ in the wave vector $\bm{k}$ space is described by the kinetic equation~\cite{Dyakonov71,Meier84:OO,Gridnev01}
\begin{equation}\label{eq:kinetic}
\frac{\partial \bm{s_k}}{\partial t} + \bm{s_k} \times \bm{\Omega_k} = \bm{g} + {\rm St} \, \bm{s_k} \:, 
\end{equation}
where $\bm{\Omega_k}$ is the Larmor frequency corresponding to the effective magnetic field, $\bm{g}$ is the spin generation rate,
e.g., due to optical excitation with circularly polarized light, and ${\rm St} \, \bm{s_k}$ is the collision integral. We consider $n$-doped QW structure with a degenerate two-dimensional electron gas 
and assume that 
optical excitation creates spin polarized electrons directly at the Fermi level, i.e., $\bm{g} \propto \delta(\varepsilon_{\bm{k}} - \varepsilon_F)$, where $\varepsilon_{\bm{k}} = \hbar^2 \bm{k}^2/(2m^*)$ is the electron kinetic energy, $m^*$ is the effective mass, and $\varepsilon_F$ is the Fermi energy. Such resonant excitation is commonly used in experiments to minimize electron gas heating~\cite{Brand02,Leyland07,Griesbeck09,Belkov08,Volkl11}. Under these conditions, the spin dephasing is determined by the effective magnetic field and details of electron scattering at the Fermi level, and energy relaxation processes are negligible.
For elastic spin-conserving scattering, the collision integral has the form~\cite{SturmanFridkin92}
\begin{equation}\label{eq:collision_int}
{\rm St} \, \bm{s_k} = \sum_{\bm{k}'} (W_{\bm{k} \bm{k}'} \bm{s}_{\bm{k}'} - W_{\bm{k}' \bm{k}}\bm{s_k}) \:,
\end{equation}
where $W_{\bm{k} \bm{k}'}$ is the rate of electron scattering from the state $\bm{k}'$ into the state $\bm{k}$ and it is assumed that the spin-orbit splitting $\hbar \Omega_{\bm{k}}$ is much smaller than the Fermi energy~\cite{Ivchenko90}. Below, we take the scattering rate in the form  $W_{\bm{k} \bm{k}'} = 2\pi \hbar^2 /(m^* L^2) \, w_{\bm{k} \bm{k}'} \, \delta(\varepsilon_{\bm{k}} - \varepsilon_{\bm{k}'})$ with 
$L^2$ being the normalization area.  

To solve Eq.~(\ref{eq:kinetic}) we decompose the distribution function $\bm{s_k}$, the frequency $\bm{\Omega_k}$, and the scattering rate $w_{\bm{k} \bm{k}'}$ into angular harmonics~\cite{Meier84:OO}
\begin{eqnarray}
\bm{s_k} &=& \sum_n \bm{s}_n {\rm e}^{i n \varphi} \:, \nonumber \\
\bm{\Omega_k}&=& \sum_{n=\pm 1} \bm{\Omega}_{n} {\rm e}^{ i n \varphi} \:, \nonumber\\
w_{\bm{k} \bm{k}'} &=& \sum_{n, m} w_{n, m} \, {\rm e}^{ i n \varphi + i m \varphi'} \:, \label{eq:harmonic_decomp}
\end{eqnarray}
where $\varphi = \arctan (k_y / k_x)$ and $\varphi' = \arctan (k'_y / k'_x)$ are the polar angles of $\bm{k}$ and $\bm{k}'$, respectively. The dominant contribution to the effective magnetic field in quantum wells is linear in the wave vector~\cite{Dyakonov86,Cartoixa06}. Therefore, we assume that the frequency $\bm{\Omega_k}$ contains only terms with $n = \pm1$; the coefficients $\bm{\Omega}_{\pm 1}$ are related by $\bm{\Omega}_{1}=\bm{\Omega}_{-1}^*$. 
The coefficients $w_{n, m}$ satisfy the relations $w_{n, m}=w_{-n, -m}^*$, $w_{n, m}=(-1)^{n+m} w_{m, n}$, and $w_{n, 0}=w_{0, n}=0$ ($n\neq 0$) which follow from reality of the scattering rate, time inversion symmetry, and the optical theorem, respectively. 
Substituting the series~(\ref{eq:harmonic_decomp}) for $\bm{s_k}$, $\bm{\Omega_k}$, and $w_{\bm{k} \bm{k}'}$ in Eq.~(\ref{eq:kinetic}) we obtain the system of linear differential equations for the angular harmonics $\bm{s}_n$
\begin{equation}\label{eq:kinetic_decomp}
\frac{d\bm{s}_n}{dt} + \sum_{m=\pm 1} \bm{s}_{n-m} \times \bm\Omega_m  =\bm{g} \, \delta_{n,0}  - w_{0,0} \bm{s}_n + \sum_m w_{n,-m} \bm{s}_m \:.
\end{equation}
Here, it is assumed that $\bm{g}$ is independent of the direction of $\bm{k}$ and, therefore, contains only zero angular harmonic. 
By solving Eqs.~(\ref{eq:kinetic_decomp}) numerically or analytically one can find the time dependence of $\bm{s}_0$ and thereby the evolution of the total spin density
$\bm{S}=(1/L^2) \sum_{\bm k} \bm{s_k} = m^*/(2\pi\hbar^2) \int_{0}^{\infty} \bm{s}_0 d\varepsilon$. 

\section{Collision-dominated regime}\label{Sec_collisions}

In this section, we consider the case of frequent electron collisions, when the spin rotation angle between scattering events is small. In this regime, the anisotropic part of the spin distribution function $\bm{s_k}$ is much smaller than $\bm{s}_0$ and Eqs.~(\ref{eq:kinetic_decomp}) can be solved iteratively~\cite{Dyakonov71,Dyakonov86}.
Such a procedure gives the following equation for the spin density
\begin{equation}\label{eq:zero_s}
\frac{d\bm{S}}{dt} = \bm{G} - \bm{\Gamma} \bm{S} \:, 
\end{equation}
where $\bm{G}=(1/L^2) \sum_{\bm{k}} \bm{g}$ is the total spin generation rate per unit area and $\bm{\Gamma}$ is the spin-relaxation-rate tensor. The latter is defined by 
\begin{equation}\label{eq:gamma_sk}
\bm{\Gamma} \bm{S} = (1/L^2) \sum_{\bm{k}} (\bm{s}_{-1} \times \bm{\Omega}_{1} + \bm{s}_{1} \times \bm{\Omega}_{-1} ) \:,
\end{equation}
where, to first oder in the effective magnetic field, the harmonics $\bm{s}_{\pm 1}$ are to be found from the equation
\begin{equation}\label{eq:delta_s}
\bm{s}_0 \times \bm{\Omega_k} = {\rm St\,} \bm{s_k} \:.
\end{equation}

The calculation of Eqs.~(\ref{eq:gamma_sk}) and~(\ref{eq:delta_s}) is similar to the calculation of an electric current density $\bm{j}$ induced by a static electric field $\bm{E}$. Indeed, the current density is expressed via the electron distribution function $f_{\bm{k}}$ by $\bm{j}=(e / L^2) \sum_{\bm{k}} (\bm{v}_1 f_{-1} + \bm{v}_{-1} f_{1})$, where $e$ is the electron charge, $\bm{v}_{\pm 1}$ and $f_{\pm 1}$ are the angular harmonics of the electron velocity $\bm{v_k}= \hbar \bm{k} / m^*$ and the distribution function, respectively. Within linear in $\bm{E}$ regime, the harmonics $f_{\pm 1}$ are found from the equation $e (df_0 / d\varepsilon_{\bm{k}}) \, \bm{v_k} \cdot \bm{E} = {\rm St} f_{\bm{k}}$, which is similar to Eq.~(\ref{eq:delta_s}). Such an analogy allows us to express the spin-relaxation-rate tensor in terms of the tensor $2 \times 2$ of in-plane electric conductivity $\bm{\sigma}$ as follows
\begin{equation}\label{eq:gamma}
\tG = \frac{\pi m^*}{e^2} \left[\bm{I}_3 \Tr(\tL \ts\trans{\tL}) - \tL\ts\trans{\tL}\right] \:,
\end{equation}
where $\bm{I}_3$ is the unit matrix $3 \times 3$, $\tL$ is the matrix $3\times 2$ relating components of the frequency $\bm{\Omega_k}$ and the wave vector $\bm{k}$,
$\bm{\Omega_k}=\tL \bm{k}$, and we used that $\ts = \ts^T$. Equations~(\ref{eq:zero_s}) and~(\ref{eq:gamma}) describe the spin dynamics of a degenerate two-dimensional electron gas for arbitrary elastic scattering
and generalize previous results. If the scattering potential is central, then the conductivity tensor is diagonal and can be expressed via the momentum relaxation time $\tau_1$ at the Fermi energy by $\ts = \sigma \bm{I}_2$, where $\sigma = \tau_1 e^2 k_F^2 /(2 \pi m^*)$, $k_F$ is the Fermi wave vector, and $\bm{I}_2$ is the unit matrix $2 \times 2$. In this particular case, Eq.~(\ref{eq:gamma}) has the form $\tG = (\tau_1 k_F^2 /2) \left[\bm{I}_3 \Tr(\tL \trans{\tL}) - \tL\trans{\tL}\right]$, in agreement with the result of D'yakonov and Kachorovskii~\cite{Dyakonov86}. 

To analyze Eq.~(\ref{eq:gamma}) in more detail we consider QW grown along $z \parallel [001]$ crystallographic direction. In such structures, the matrix $\tL$ has nonzero components
\begin{equation}\label{eq:lambda}
\Lambda_{xy}=\alpha+\beta \:, \;\; \Lambda_{yx}=\beta-\alpha \:,
\end{equation}
where $\alpha$ and $\beta$ are the constants of the Rashba and Dresselhaus spin-orbit splitting, respectively, $x\parallel [1\bar{1} 0]$ and $y\parallel [110]$ are the in-plane axes~\cite{Averkiev99,Averkiev08}. Then, components of the tensor $\tG$ take the form
\begin{eqnarray}\label{eq:gamma_components}
\Gamma_{xx}= \frac{\pi m^*}{e^2} (\alpha - \beta)^2 \sigma_{xx} \:, \;\; \Gamma_{yy} = \frac{\pi m^*}{e^2} (\alpha + \beta)^2 \sigma_{yy} \:, \nonumber \\ 
\Gamma_{xy}=\Gamma_{yx}=\frac{\pi m^*}{e^2} (\alpha^2 - \beta^2) \sigma_{xy} \:, \;\; \Gamma_{zz}=\Gamma_{xx}+\Gamma_{yy} \:. \;\;\;
\end{eqnarray}

\begin{figure}[b]
 \includegraphics[width=0.9\linewidth]{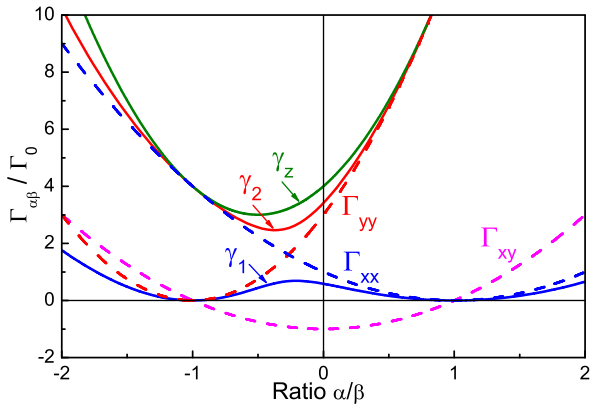}
 \caption{(Color online) Dependences of the spin-relaxation-rate tensor components $\Gamma_{xx}$, $\Gamma_{yy}$, and $\Gamma_{xy}$ (dashed curves) and the tensor eigen values $\gamma_1$, $\gamma_2$, and $\gamma_z=\Gamma_{zz}$ (solid curves) on the ratio $\alpha / \beta$  calculated for $\sigma_{xx} / \Tr \ts = \sigma_{xy} / \Tr \ts =1/4$. The curves are normalized by $\Gamma_0 = (\pi m^* /e^2) \, \beta^2 \Tr \ts$.}
 \label{figure1}
\end{figure}

Dependences of the tensor components $\Gamma_{xx}$, $\Gamma_{yy}$, and $\Gamma_{xy}$ on the ratio $\alpha / \beta$ for QWs with strong scattering anisotropy are plotted in Fig.~1 by dashed curves. 
The in-plane eigen values $\gamma_1$ and $\gamma_2$ of the tensor $\tG$ and the out-of-plane value $\gamma_z$, which coincides with $\Gamma_{zz}$, are found from the equation ${\rm det} (\gamma \bm{I}_3 -  \tG)=0$ and presented in Fig.~\ref{figure1} by solid curves. One can see that the minimum of $\gamma_z$, which corresponds to the longest lifetime of the spin component $S_z$, is achieved in asymmetric QWs where the Rashba constant $\alpha \neq 0$. This is in contrast to QW structures with central electron scattering where the Rashba splitting is known to decrease the spin lifetime. The analysis of Eqs.~(\ref{eq:gamma_components}) shows that, at fixed $\beta$, the rate $\gamma_z$ reaches the minimum at $\alpha / \beta = (\sigma_{xx}-\sigma_{yy})/ \Tr \ts$. We also note that $\gamma_1 \neq \gamma_2$ no matter how small the ratio $\alpha/ \beta$ is if the eigen axes of the conductivity tensor do not coincide with $x$ and $y$, i.e., $\sigma_{xy} \neq 0$. 

Now we consider the effect of an external magnetic field $\bm{B}$ on spin dephasing. The magnetic field causes the Larmor precession of electron spins and cyclotron motion of electrons in QW plane with the frequencies $\bm{\Omega}_L = g \mu_B \bm{B} / \hbar$ and $\omega_c = e B_z / (m^* c)$, respectively~\cite{Ivchenko73,Wilamowski04}. Here, $g$ is the effective electron $g$-factor, $\mu_B$ is the Bohr magneton, $e$ is the electron charge, and $c$ is the speed of light. Both effects are theoretically described in the framework of kinetic approach, with kinetic equation having the form~\cite{Ivchenko73,Glazov04,Tarasenko09}
\begin{equation}\label{kinetic_field}
\frac{\partial \bm{s_k}}{\partial t} + \bm{s_k} \times (\bm{\Omega_k} + \bm{\Omega}_L) - \omega_c \left[ \bm{k} \times \frac{\partial}{\partial\bm{k}} \right]_z \bm{s_k} = \bm{g} + {\rm St} \, \bm{s_k} \:. 
\end{equation}

Solution of Eq.~(\ref{kinetic_field}) shows that, in the collision-dominated regime, the time evolution of the spin density $\bm{S}$ is described by Eqs.~(\ref{eq:zero_s}) and~(\ref{eq:gamma}) where (i) the additional term $\bm{S} \times \bm{\Omega}_L$ is added to the left-hand side of Eq.~(\ref{eq:zero_s}) and (ii) $\ts$ in Eq.~(\ref{eq:gamma}) is replaced by the transposed tensor of electric conductivity in the magnetic field $\ts^T(\omega_c)$.
We note that in zero magnetic field the conductivity tensor is symmetric, i.e., $\ts^T(0)=\ts(0)$. In the presence of magnetic field, the tensor $\ts(\omega_c)$ contains both  symmetric $\smt{\ts}(\omega_c)=[{\ts}(\omega_c) + \trans{\ts}(\omega_c)] / 2$ and antisymmetric $\asmt{\ts}=[{\ts}(\omega_c) - \trans{\ts}(\omega_c)] / 2$ parts. Accordingly, the right-hand side of Eq.~(\ref{eq:gamma}) can be also reduced to the sum of symmetric $\tG^{(s)}(\omega_c)$ and antisymmetric $\tG^{(a)}(\omega_c)$ tensors. The symmetric tensor $\tG^{(s)}(\omega_c)$ describes spin relaxation. The antisymmetric third-rank tensor $\tG^{(a)}(\omega_c)$ is equivalent
to a pseudovector $\delta\bm{\Omega}_L$ and represents, in fact, a correction to the Larmor frequency~\cite{Ivchenko73,Edelstein06,Tarasenko09}. Therefore, the equation describing the time evolution of spin density in the magnetic field has the final form
\begin{equation}\label{eq:spineq_b}
\frac{d\bm{S}}{dt} + \bm{S} \times (\bm{\Omega}_L + \delta \bm{\Omega}_L ) = \bm{g} - \tG(\omega_c) \bm{S} \:,
\end{equation}
where
\begin{equation}\label{eq:gamma_field}
\tG(\omega_c) = \frac{\pi m^*}{e^2} \left\{ \bm{I}_3 \Tr\left[\tL\smt{\ts}(\omega_c)\trans{\tL}\right] - \tL\smt{\ts}(\omega_c)\trans{\tL}\right\} \\
\end{equation}
is the spin-relaxation-rate tensor,
\begin{equation}\label{eq:omega_b}
{\left( \delta\bm{\Omega}_L \right)}_\alpha = \frac{\pi m^*}{2 e^2} \sum_{\beta\gamma} {\epsilon}_{\alpha \beta \gamma} \left[\tL\asmt{\ts}(\omega_c)\trans{\tL}\right]_{\beta \gamma} 
\end{equation}
is the Larmor frequency correction caused by cyclotron motion, and ${\epsilon}_{\alpha \beta \gamma}$ is the antisymmetric third-rank tensor (Levi-Civita symbol).

\begin{figure}[ht]
\includegraphics[width=0.85\linewidth]{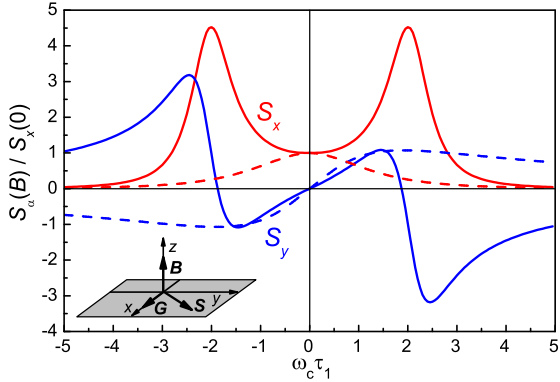}
\caption{(Color online) Magnetic field dependences of the spin components $S_x$ and $S_y$ calculated for $\bm{G} \parallel x$, $\bm{B} \parallel z$, the Rashba spin-orbit splitting, $\alpha k_F \tau_1 = 0.3$, and $\Omega_L / \omega_c = - 0.01$ (solid curves) or $\Omega_L / \omega_c = + 0.01$ (dashed curves).}
 \label{figure2}
\end{figure}

As follows from Eq.~(\ref{eq:spineq_b}), the precession of total electron spin is determined by the frequency $\bm{\Omega}_L + \delta \bm{\Omega}_L$. The frequency correction $\delta \bm{\Omega}_L$ depends on the magnetic field non-monotonically: It is proportional to the magnetic field at small fields, reaches maximum at $\omega_c \tau_1 \sim 1$, and decreases with the further field increase. It may happen that $\bm{\Omega}_L$ and $\delta \bm{\Omega}_L$ have opposite signs and compensate each other at a certain magnetic field. Such a compensation results in a peculiarity in the magnetic field dependence of the electron spin. As an example, we consider the simple case of continuous spin generation, central electron scattering, and the magnetic field $\bm{B}$ pointed along the QW normal $[001]$. Then, the conductivity-tensor components have the form $\sigma_{xx}(\omega_c) = \sigma_{yy}(\omega_c) = \sigma/[1+(\omega_c\tau_1)^2]$, $\sigma_{xy}(\omega_c) = - \sigma_{yx}(\omega_c)= \sigma \omega_c\tau_1/[1+(\omega_c\tau_1)^2]$, and the spin-relaxation-rate tensor (\ref{eq:gamma_field}) is diagonal in the chosen coordinate frame $(x,y,z)$. Straightforward calculation shows that the components of the steady-state spin density $\bm{S}$ have the form
\begin{eqnarray}\label{eq:static}
S_x &=& \frac{\Gamma_{yy}(\omega_c) \, G_x - (\Omega_L + \delta \Omega_L) G_y}{\Gamma_{xx}(\omega_c)\, \Gamma_{yy}(\omega_c) + (\Omega_L + \delta \Omega_L)^2} \:, \nonumber \\
S_y &=& \frac{\Gamma_{xx}(\omega_c) \, G_y + (\Omega_L + \delta \Omega_L) G_x}{\Gamma_{xx}(\omega_c)\, \Gamma_{yy}(\omega_c) + (\Omega_L + \delta \Omega_L)^2} \:, \nonumber \\
S_z &=& \frac{G_z}{\Gamma_{zz}(\omega_c)} \:, 
\end{eqnarray}
where $\tG(\omega_c)= \tG(0) / [1+(\omega_c \tau_1)^2]$, $\Gamma_{xx} = \tau_1 k_F^2 (\alpha-\beta)^2/2$, $\Gamma_{yy} = \tau_1 k_F^2 (\alpha+\beta)^2/2$, $\Gamma_{zz} = \Gamma_{xx} + \Gamma_{yy}$~\cite{Averkiev99}, and 
\begin{equation}\label{eq:delta_simple}
\delta \Omega_L = \frac{ \tau_1 k_F^2(\alpha^2 - \beta^2)}{2} \frac{\omega_c \tau_1}{1+(\omega_c \tau_1)^2} \:.
\end{equation}
Shown in Fig.~\ref{figure2} are the magnetic field dependences of the in-plane components $S_x$ and $S_y$ calculated for the spin generation $\bm{G} \parallel x$, the Rashba spin-orbit splitting, and $\Omega_L / \omega_c = \pm 0.01$. Such ratios of the Larmor to cyclotron frequency can be realized, e.g., in GaAs/AlGaAs QW structures~\cite{Kiselev98}. The dependences plotted for $\Omega_L / \omega_c = 0.01$ (dashed curves) are rather simple: 
$S_x$ is maximal at $B=0$ and monotonically decays with the field increase; $S_y \propto B$ at small fields, reaches maximum, and then decays. In contrast, the magnetic field dependences of $S_x$ and $S_y$ calculated for $\Omega_L / \omega_c = - 0.01$ (solid curves) are completely different. The component $S_x$ reaches maximum at a finite magnetic field  corresponding to $\omega_c\tau_1 \approx 2$. $S_y$ has two extrema for a fixed direction of $\bm{B}$ and changes the sign approximately at the magnetic field where $S_x$ is maximal. Such a behavior is caused by interference of $\Omega_L$ and $\delta\Omega_L$ which compensate each other at $\omega_c \tau_1 \approx 2$ for the parameters chosen. The vanishing of the total Larmor frequency leads to the Hanle-like curves in the vicinity of this magnetic field, see Fig.~\ref{figure2}. The fact that $S_x$ in the point of compensation is much larger than $S_x(0)$ is caused by a slowdown of the DP spin dephasing by cyclotron motion~\cite{Ivchenko73}, see Eq.~(\ref{eq:static}). 
%

\section{Oscillatory regime}\label{Sec_oscillations}

Now we consider the oscillatory regime of spin dephasing which occurs if the relaxation time $\tau_1$ is longer than $1/\Omega_{\bm{k}}$~\cite{Gridnev01,Brand02,Leyland07,Griesbeck09}. For arbitrary $\bm{\Omega}_{\bm{k}}$ and scattering rate $w(\varphi,\varphi')$, Eqs.~(\ref{eq:kinetic_decomp}) can be solved only numerically. Therefore, we focus on scattering anisotropy and assume the isotropic spin-orbit splitting of the Rashba type. 
The scattering rate can be conveniently presented as the sum of two terms~\cite{SturmanFridkin92} 
\begin{equation}
w(\varphi, \varphi') = \cen{w}(|\varphi - \varphi'|) +  \delta w(\varphi, \varphi') \:,
\end{equation}
where $\cen{w}(\phi) =  \int_0^{2\pi} w(\varphi'+\phi,\varphi')\, d\varphi'/(2\pi)$, $\delta w(\varphi, \varphi')=w(\varphi, \varphi')-\cen{w}(|\varphi - \varphi'|)$, and we assume that $\delta w \ll \cen{w}$. The term $\cen{w}(\phi)$ describes central scattering. The corresponding collision integral is expressed in terms of the relaxation times $\tau_n$ of angular harmonics of the distribution function, 
$1/\tau_n = \int_0^{2\pi} \cen{w}(\phi) (1-\cos n\phi) \: d\phi/(2\pi)$. Then, Eqs.~(\ref{eq:kinetic_decomp}) take the form
\begin{eqnarray}
\frac{ds_{z,n}}{dt} - \Omega_R \frac{s_{-,n-1} + s_{+,n+1}}{2} = g_z \delta_{n,0} - \frac{s_{z,n}}{\tau_n} \hspace{1cm} \nonumber \\
+  \sum_m \delta w_{n,-m} \, s_{z,m} \:,\;\; \label{eq:osc_decomp_z} \\
\frac{ds_{\pm,n}}{dt}+\Omega_R \, s_{z,n \mp 1} =  g_{\pm} \delta_{n,0} - \frac{s_{\pm,n}}{\tau_n} \hspace{2.3cm} \nonumber \\
+ \sum_m \delta w_{n,-m} \, s_{\pm,m}  \,,\;\; \label{eq:osc_decomp_pm}
\end{eqnarray}
where $s_{\pm,n} = s_{x,n} \pm i s_{y,n}$, $g_{\pm} = g_{x} \pm i g_{y}$, and $\Omega_R = \alpha k_F$ is the precession frequency corresponding to the Rashba field at the Fermi level.

In the regime of continuous spin generation, when $\bm{g}$ is independent of time, Eqs.~(\ref{eq:osc_decomp_z}) and~(\ref{eq:osc_decomp_pm}) can be solved iteratively.  
To first oder in $\delta w$, spin components depend on the harmonics $\delta w_{11}=\delta w_{-1,-1}^*$ and $\delta w_{2,-1}=\delta w_{-2,1}^*=-\delta w_{-1,2}=-\delta w_{1,-2}^*$. Such angular harmonics in the scattering rate reduce space symmetry of the system and, in fact, correspond to a symmetric tensor and an in-plane vector, respectively. Accordingly, we define $2 \times 2$ tensor $\bm{Q}$ by $Q_{xx} = -Q_{yy} = \Re \delta w_{1,1}$ and $Q_{xy} = Q_{yx} = -\Im \delta w_{1,1}$ and the vector $\bm{\nu}$ by $\nu_x = \Re \delta w_{2,-1}$ and $\nu_y = - \Im \delta w_{2,-1}$.
In these notations, the in-plane $\bm{S}_\parallel=(S_x,S_y)$ and out-of-plane $S_z$ components of the steady-state spin density are given by 
\begin{eqnarray}
\bm{S}_\parallel &=& \left(\tau_2 + \frac{2}{\Omega_R^2 \tau_1}\right) \bm{G}_\parallel - \frac{2}{\Omega_R^2} \bm{Q}  \bm{G}_\parallel + \frac{\tau_2 \, \bm{\nu}}{\Omega_R} G_z \:,\;\; \label{eq:osc_spin_p} \\
S_z &=& \frac{G_z}{\Omega_R^2 \tau_1} + \frac{\tau_2 \, \bm{\nu} \cdot \bm{G}_\parallel}{\Omega_R} \:, \label{eq:osc_spin} 
\end{eqnarray}
where $\bm{G}_\parallel=(G_x,G_y)$ is the projection of generation rate onto the QW plane. Equations~(\ref{eq:osc_spin_p}) and~(\ref{eq:osc_spin}) show that the in-plane and out-of plane spin components are coupled in structures with anisotropic scattering; the coupling strength is proportional to $\bm{\nu}$. In particular, the generation of electron spin along the QW normal, i.e., 
$\bm{G} \parallel z$, 
leads not only to $S_z$ but also to $\bm{S}_\parallel \propto \bm{\nu} G_z$. Moreover, even in the case of small scattering anisotropy, the in-plane and out-of-plane spin components can be comparable to each other provided $\Omega_R \tau_1$ is large enough. The second term on the right-hand side of Eq.~(\ref{eq:osc_spin_p}) describes the in-plane anisotropy of spin dephasing due to anisotropic conductivity which, to first order in $\delta w$, has the form $\bm{\sigma}=\tau_1 e^2 k_F^2 /(2\pi m^*) (\bm{I}_2 + \tau_1 \bm{Q})$.
The effect of conductivity anisotropy on spin relaxation was considered in Sec.~\ref{Sec_collisions}. Below we focus on the coupling between $\bm{S}_\parallel$ and $S_z$ and assume, for simplicity, that $\bm{Q}=0$. 

\begin{figure}[t]
 \includegraphics[width=0.85\linewidth]{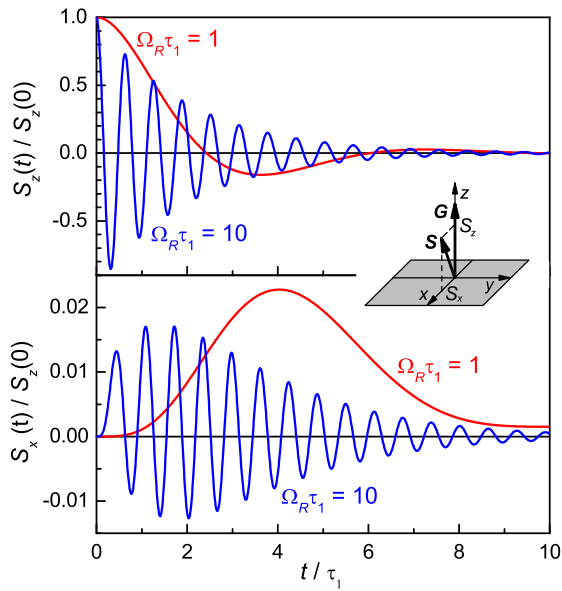}
 \caption{(Color online) Time dependences $S_z(t)$ and $S_x(t)$ after a short optical pulse orienting electron spins along $z$. The curves are calculated for $\bm\nu \parallel x$, $\nu\tau_1=0.1$, $\tau_n =\tau_1$, and two different $\Omega_R \tau_1$.}
 \label{figure3}
\end{figure}

The coupling between the in-plane and out-of-plane components of the spin density 
can be also studied in experiments with high time resolution. Shown in Fig.~\ref{figure3} are the time dependences $S_z(t)$ and $S_x(t)$ after a short circularly polarized optical pulse which orients electron spins along $z$ at $t=0$. The curves are obtained by solving Eqs.~(\ref{eq:osc_spin_p}) and~(\ref{eq:osc_spin}) numerically for noncentrosymmetric scattering potential with $\bm{\nu} \parallel x$. One can see that $S_z(t)$ demonstrates damping oscillations
as is expected for the oscillatory regime of spin dephasing~\cite{Gridnev01}. The oscillations are caused by precession of individual electron spins in the effective magnetic field.
The in-plane spin component $S_x$ is zero right after the pulse, emerges at the time scale of momentum relaxation, and then also decays. For the parameters given in caption to Fig.~\ref{figure3}, $S_x(t)$ reaches a few percent of $S_z(0)$. 

\begin{figure}[t]
 \includegraphics[width=0.99\linewidth]{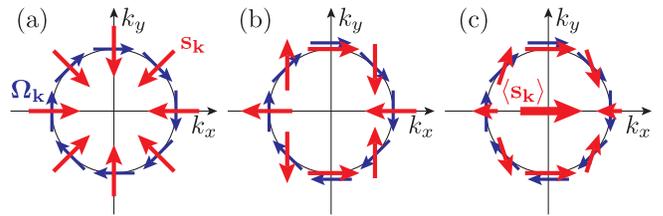}
 \caption{(Color online) Microscopic mechanism of the generation of in-plane spin polarization when electron spins are initially oriented along the QW normal. Precession of electron spins in the effective magnetic field followed by anisotropic electron scattering and subsequent spin precession in the effective field results in a spin polarization along $x$.}
 \label{figure4}
\end{figure}

Microscopic mechanism of the generation of the in-plane component $S_x$ is a three-stage process illustrated in Fig.~\ref{figure4}. At the first stage [Fig.~\ref{figure4}(a)], electron spins initially oriented along $z$ precess in the Rashba field with the frequency $\bm{\Omega_k}$. The precession forms a spin distribution function $\bm{s}_{\bm{k}}$ containing the first angular harmonic. The electron scattering by noncentrosymmetric defects modifies $\bm{s}_{\bm{k}}$ and, due to the terms $\propto \delta w_{2,-1}$ and $\propto \delta w_{-2,1}$ in the collision integral, partially transforms the first angular harmonic into the second harmonic. The spin distribution function described by the second angular harmonic is shown in Fig.~\ref{figure4}(b). Finally [Fig.~\ref{figure4}(c)], the subsequent rotation of electron spins in the Rashba field results in a net spin polarization of carriers along the $x$ axis. 
\begin{figure}[t]
 \includegraphics[width=0.95\linewidth]{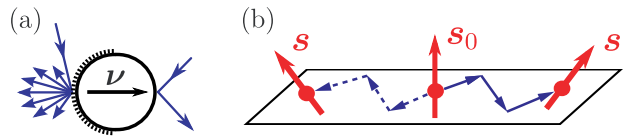}
 \caption{(Color online) (a) Example of a noncentrosymmetric scatterer. Disk with different edges which diffusively and specularly scatter electrons. (b) Electron trajectories in QW plane. Trajectories shown by solid and dashed lines are connected by space inversion.}
 \label{figure5}
\end{figure}

The coupling between $\bm{S}_{\parallel}$ and $S_z$ can be also understood by analyzing electron trajectories in QW structures where the angular dependence of scattering rate contains 
the harmonics $\delta w_{2,-1}$ and $\delta w_{-2,1}$.
An example of scatterer providing such harmonics is a disk, one edge of which reflects electrons specularly while the other one scatters electrons diffusively, see Fig.~\ref{figure5}(a). Obviously, such scatterers with a preferred orientation in the QW plane
break the in-plane space inversion~\footnote{Together with inversion asymmetry along the $z$ axis causing the Rashba splitting, the scatterers reduce the overall point group of the structure to $C_s$.}. 
The scattering anisotropy modify electron trajectories, which affects the spin dynamics.  
Indeed, in QWs with centrosymmetric scatterers, each electron trajectory has on average its counterpart connected by space inversion, see solid and dashed lines in Fig.~\ref{figure5}(b). Electrons the spin $\bm{s}_0$ initially oriented along $z$ move in the QW plane and gain in-plane spin components $\bm{s}_{\parallel}$ due to rotation in the Rashba field. However, the particles propagating along the paths interconnected by space inversion gain the opposite projections $\bm{s}_{\parallel}$ leading to a vanishing average in-plane spin polarization. In QWs with noncentrosymmetric scatterers, the space-inversion symmetry of electron trajectories is broken, which results in a non-zero in-plane spin polarization. To first order in scattering anisotropy, $\bm{S}_{\parallel}$ is determined by the angular harmonics $\delta w_{2,-1}$ and $\delta w_{-2,1}$. Other harmonics $\delta w_{n,m}$ describing noncentrosymmetric scattering
can also couple the in-plane and out-of-plane spin components in higher orders in $\delta w$. 

Equations~(\ref{eq:osc_decomp_z})-(\ref{eq:osc_spin}) are obtained for the Rashba spin-orbit splitting from general Eq.~(\ref{eq:kinetic_decomp}). Similar calculations can be carried out for the case of Dresselhaus splitting. One can see that Eq.~(\ref{eq:kinetic_decomp}) is invariant to the replacement of the Rashba field with the Dresselhaus one, which differ in sign of $\Omega_y$, and the simultaneous inversion of $s_x$ and $g_x$ signs. Thus, Eqs.~(\ref{eq:osc_spin_p}) and~(\ref{eq:osc_spin}) where $\Omega_R$, $S_x$, and $G_x$ are replaced by $\Omega_D = \beta k_F$, $-S_x$, and $-G_x$, respectively, describe the steady-state spin density in (001)-grown QWs with the Dresselhaus splitting. In general case, if both the Rashba and Dresselhaus contributions to the effective field are present, a prerequisite for the coupling between $\bm{S}_{\parallel}$ and $S_z$ remains the lack of inversion symmetry in scattering potential. We also note that the coupling is absent irrespective of the form of $w(\varphi,\varphi')$ if $|\alpha|=|\beta|$ because, in this particular case, the frequency $\bm{\Omega}_{\bm{k}}$ depends only upon one component of the wave vector.      

Finally, we discuss the effect of an external magnetic field on spin dynamics in the oscillatory regime. Equations~(\ref{eq:osc_decomp_z}) and~(\ref{eq:osc_decomp_pm}) with the Larmor precession and cyclotron motion being taken into account have the form
\begin{eqnarray}
\frac{ds_{z,n}}{dt} - \Omega_R \frac{s_{+,n+1} + s_{-,n-1}}{2} + i \frac{\Omega_{L,-} s_{+,n} - \Omega_{L,+} s_{-,n}}{2}  \nonumber \\ = g_z \delta_{n,0} - \left( \frac{1}{\tau_n} - i n \omega_c \right) s_{z,n} +  \sum_m \delta w_{n,-m} \, s_{z,m} \:,\;\;\; \label{eq:osc_B1} \\
\frac{ds_{\pm,n}}{dt}+\Omega_R \, s_{z,n \mp 1} \pm i \Omega_{L,\pm} \, s_{z,n} =  g_{\pm} \delta_{n,0}  \hspace{2cm} \nonumber \\ - \left( \frac{1}{\tau_n} - i n \omega_c \mp i \Omega_{L,z} \right) s_{\pm,n} + \sum_m \delta w_{n,-m} \, s_{\pm,m}  \,,\;\;\; \label{eq:osc_B2}
\end{eqnarray}
where $\Omega_{L,\pm} = \Omega_{L,x} \pm i \Omega_{L,y}$. Equations~(\ref{eq:osc_B1}) and~(\ref{eq:osc_B2}) are valid for arbitrary strength of spin-orbit splitting $\Omega_R \tau_n$ and angular dependence of the scattering rate.

\begin{figure}[b]
 \includegraphics[width=0.99\linewidth]{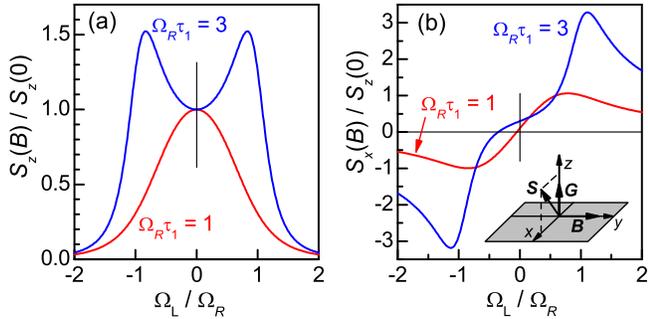}
 \caption{(Color online) Dependences $S_z$ and $S_x$ on the in-plane magnetic field $\bm{B} \parallel y$ measured in units of $\Omega_L/\Omega_R$. The curves are calculated for $\bm{G} \parallel z$, $\bm\nu \parallel x$, $\tau_n =\tau_1$, $\nu\tau_1=0.1$, and two different $\Omega_R \tau_1$.}
 \label{figure6}
\end{figure}
Dependences of the steady-state components $S_z$ and $S_x$ on the in-plane magnetic field $\bm{B}\parallel y$ for continuous spin generation along the $z$ axis are shown in Fig.~\ref{figure6}. The curves are obtained by solving Eqs.~(\ref{eq:osc_B1}) and~(\ref{eq:osc_B2}) numerically for $\bm{\nu} \parallel x$ and different $\Omega_R \tau_1$.
One can see that the curves drastically depend on the parameter $\Omega_R \tau_1$. At $\Omega_R \tau_1 =1$, the dependences $S_z(B)$ and $S_x(B)$ are similar to conventional Hanle curves. The only difference is that $S_x(0) \neq 0$ due to scattering anisotropy, see Eq.~(\ref{eq:osc_spin_p}). 
The dependences $S_z(B)$ and $S_x(B)$ calculated for large $\Omega_R \tau_1$, $\Omega_R \tau_1 =3$ in Fig.~\ref{figure6}, look completely different. Instead of a monotonic decrease with the magnetic field, $S_z(B)$ increases first with the field, reach maximum at $\Omega_L \approx \Omega_R$, and then decreases. $S_x(B)$ is nearly independent of $B$ at small magnetic fields and exhibits a sharp rise at $\Omega_L \approx \Omega_R$.
Such a behavior is caused by a partial suppression of the DP spin dephasing mechanism by the external in-plane magnetic field equal to the effective field in high-mobility structures~\cite{Poshakinskiy11}. 
We also note that the dependence $S_z(B)$ is always even despite the fact that the vector $\bm{S}(0)$ is not aligned along the $z$ axis. The evenness of $S_z(B)$ follows from Eqs.~(\ref{eq:osc_B1}) and~(\ref{eq:osc_B2}).       

\begin{figure}[b]
 \includegraphics[width=0.85\linewidth]{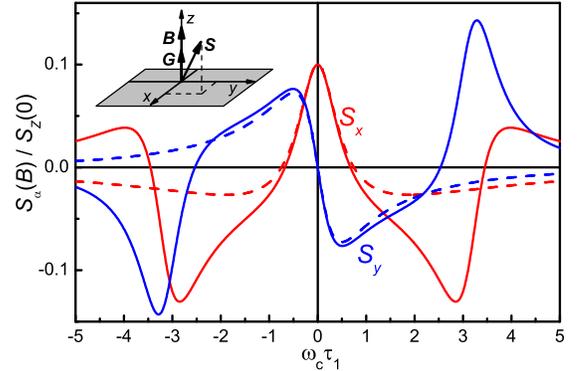}
 \caption{(Color online) Dependences $S_x$ and $S_y$ on the magnetic field $\bm{B} \parallel z$ measured in units of $\omega_c \tau_1$. The curves are calculated after Eqs.~(\ref{eq:osc_normal}) for $\bm\nu \parallel x$, $\tau_n =\tau_1$, $\nu\tau_1=0.1$, $\Omega_R \tau_1 =1$, and 
$\Omega_L/\omega_c = - 0.05$ (solid curves) or $\Omega_L/\omega_c = + 0.05$ (dashed curves).}
 \label{figure7}
\end{figure}

For the magnetic field pointed along the QW normal ($\bm{\Omega}_L \parallel z$) and continuous spin generation, the steady-state solution of Eqs.~(\ref{eq:osc_B1}) and~(\ref{eq:osc_B2}) can be found analytically. To first order in the scattering asymmetry $\delta w$ and for $\bm{G}\parallel z$, the solution has the form
\begin{eqnarray}
S_x &=& {\rm Re} \left[ \frac{(\nu_x - i\nu_y)  \,\Omega_R \, \tilde{\tau}_{1} \tilde{\tau}_{2} }{1 + i \Omega_L \tilde{\tau}_2 + 2i\Omega_L (1/\tau_1 - i\omega_c)/\Omega_R^2} \right] S_z \:, \nonumber  \\
S_y &=& {\rm Re} \left[ \frac{(\nu_y + i\nu_x)  \,\Omega_R \, \tilde{\tau}_{1} \tilde{\tau}_{2} }{1 + i \Omega_L \tilde{\tau}_2 + 2i\Omega_L (1/\tau_1 - i\omega_c)/\Omega_R^2} \right] S_z \:, \nonumber  \\
S_z &=& \frac{1+(\omega_c + \Omega_L)^2\tau_1^2}{\Omega_R^2 \tau_1} G_z \:, \label{eq:osc_normal}
\end{eqnarray}
where $1/\tilde{\tau}_n = 1/\tau_n - in\omega_c - i\Omega_L$. The $z$-component of the spin density quadratically increases with the magnetic field growth, which is caused by a slowdown of the D'yakonov-Perel' spin dephasing mechanism in the perpendicular magnetic field~\cite{Wilamowski04,Glazov04}. The dependences of $S_x$ and $S_y$ on the magnetic field $\bm{B} \parallel z$ are more complicated and drastically depend on parameters. Examples of such dependences are shown in Fig.~\ref{figure7}. First, we note that $S_x \neq 0$ in zero magnetic field due to scattering anisotropy. With the field $B$ increase, $S_x$ decreases and changes the sign. The component $S_y$ depends linearly on $B$ at small magnetic fields, reaches an extremum, and then decreases. The curves calculated for $\Omega_L/\omega_c = - 0.05$ (solid curves) have additional Hanle-like peculiarities at $\omega_c\tau_1 \approx 3.2$. The analysis of Eqs.~(\ref{eq:osc_normal}) shows that the peculiarities occur at $\omega_c \approx \Omega_R  \sqrt{-\omega_c/(2\Omega_L)}$ and are of similar origin as those in Fig.~\ref{figure2} caused by nulling the total frequency $\Omega_L + \delta\Omega_L$.
%

\section{Summary}\label{Summary}

We have developed the microscopic theory of electron spin dephasing in QW structures with anisotropic scatterers. It is shown that, in the collision-dominated regime of spin dephasing, the spin-relaxation-rate tensor is determined by constants of spin-orbit splitting and the electric conductivity tensor. In (001)-grown structures with anisotropic in-plane conductivity, the longest spin lifetime of electrons polarized along the QW normal is achieved in asymmetric QWs with a finite Rashba splitting. We have demonstrated that, in structures with
noncentrosymmetric scattering potentials, the in-plane and out-of-plane spin components are coupled to each other. The coupling is caused by breaking the space-inversion symmetry of electron trajectories in the QW plane and is more pronounced in structures with strong spin-orbit splitting. The engineering of electron trajectories provides an additional approach to manipulate electron spins in low-dimensional semiconductors. 

\paragraph*{Acknowledgments.}  This work was supported by the RFBR, Russian Ministry for Education and Science (contract 14.740.11.0892), EU programs ``Spinoptronics'' and ``POLAPHEN'', and the Foundation ``Dynasty''-ICFPM.

\bibliography{spinrel}

\end{document}